\documentclass[preprint,aps,showpacs,floats,superscriptaddress,nofootinbib]{revtex4}
\usepackage{psfig}
\usepackage[dvips]{graphicx,psfrag}

\newcommand{\lsim}{\buildrel<\over{_\sim}}
\newcommand{\gsim}{\buildrel>\over{_\sim}}

\newcommand{\stau}{\tilde{\tau}}
\newcommand{\betastau}{\beta_{\tilde{\tau}}}
\newcommand{\mstau}{m_{\tilde{\tau}}}

\newcommand{\se}{\tilde{e}}
\newcommand{\betase}{\beta_{\tilde{e}}}
\newcommand{\mse}{m_{\tilde{e}}}

\begin{document}

\preprint{YITP-04-50}
\preprint{DESY-04-179}
\title{A study of late decaying charged particles at future colliders
}
\author{Koichi Hamaguchi}
\affiliation{Deutsches Elektronen-Synchrotron DESY, 22603 Hamburg, Germany}
\author{Yoshitaka Kuno} 
\affiliation{Department of Physics, Graduate School of Science, 
Osaka University, Toyonaka, Osaka 560-0043, Japan}
\author{Tsuyoshi Nakaya }
\affiliation{Department of Physics,
Kyoto University, Kyoto , Japan}
\author{Mihoko M. Nojiri}
\affiliation{YITP, Kyoto University, Kyoto 606-8502, Japan}

\date{\today}

\begin{abstract}
In models where the gravitino is the lightest supersymmetric
particle (LSP), the next-to-lightest supersymmetric particle (NLSP) is
long-lived.  
We consider an important charged NLSP candidate, the scalar tau 
$\tilde{\tau}$. Slow charged NLSPs may be produced at future 
colliders and they may be stopped in 
a massive stopper which simultaneously serves as a detector 
for  the NLSP and 
its decay products. We found the number of events at a 1~kton to O(10)~kton 
detector could be significant enough to study the NLSP decays
with lifetime shorter than $10^{10}$ sec at the LHC.  The performance of
existing 1~kton detectors may be good enough to do such studies at the
LHC, if they can be placed close to the ATLAS/CMS detectors.  At a
future $e^- e^-$ collider, scalar electrons $\tilde{e}^-$s are
copiously produced.  Slow NLSPs may be produced from the
$\tilde{e}^-$ decay.  The number of stopped NLSPs at a future linear collider
could be large enough to study
rare decay modes of the NLSP.
\end{abstract}
\maketitle

\section{Introduction}
Supersymmetry (SUSY) is one of the key features of 
theories of quantum gravity, such as superstring. 
Supersymmetric extension of gravity (supergravity),
or local SUSY,
predicts a spin-$3/2$ particle, the gravitino. 
When supersymmetry is spontaneously broken, 
the gravitino acquires a
mass by absorbing a massless goldstino.
In most of the supersymmetric models, SUSY is broken 
in a hidden sector.
While the masses of superpartners of the standard model (SM) particles 
may depend on the mechanism to mediate SUSY breaking in the 
hidden sector, 
the gravitino mass is only described by the 
SUSY breaking scale, $F$, and the Planck scale, $M_{\rm P}$, 
as $m_{3/2}=F/(\sqrt{3}M_{\rm P})$.
Discovery of the gravitino would be therefore one of the most 
direct ways to prove the nature of the hidden sector 
of the supersymmetric models. 

The rate of the direct production of a gravitino from particle
collisions is small.  If the gravitino is the lightest supersymmetric
particle (LSP), it is produced from cascade decays of heavier
sparticles, and its nature can be thus studied.  The LSP gravitino is
also cosmologically important.  The reheating temperature after
inflation is severely constrained by the `overclosure'
problem \cite{Moroi:1993mb}. There have been, however, proposed
several mechanisms which avoid this constraint and which, in addition,
explain the cold dark matter density in terms of thermally produced
LSP gravitinos \cite{gravCDM}.  Late time decays of 
next-to-lightest supersymmetric particles (NLSPs)
might change the
predictions of big bang nucleosynthesis (BBN) if the lifetime
is longer than O(1) sec \cite{recentBBN}, which leads to
constraints on the masses of NLSP
and gravitino \cite{Feng:2004mt}, 
and parameters of supersymmetric models \cite{gravLSPinCMSSM}.
Gravitino dark matter scenario may also be realized non-thermally, e.g.
via NLSP decays \cite{gravSWIMP} 
or right-handed sneutrino decay \cite{Allahverdi:2004ds}.

In the LSP gravitino scenario, the NLSP is metastable, and they escape
detector before decaying into the gravitino.
The NLSP is likely the
scalar tau lepton $\tilde{\tau}$ (stau), because the slepton masses
receive smaller radiative corrections from gaugino masses 
than 
squarks, and the stau mass receives a larger 
Yukawa correction than
the other sleptons.  The lighter mass eigenstate $\tilde{\tau}$, which
is a linear combination of $\tilde{\tau}_L$ and $\tilde{\tau}_R$,
becomes even lighter due to the left-right mixing.
The collider phenomenology has been studied in 
Refs.~[\onlinecite{Drees:1990yw,Feng:1997zr-Martin:1998vb,short},
\onlinecite{Feng:2004mt}].

The main decay mode of the NLSP stau is $\stau\to \tau\,\psi_{3/2}$, 
where $\psi_{3/2}$ is the gravitino.
The decay width  depends only on the stau mass $\mstau$,
the gravitino mass $m_{3/2}$, and the Planck scale $M_{\rm P}$. It is  
given by the following formula
\begin{equation}
\Gamma_{\tilde{\tau}}(\tilde{\tau}\to\psi_{3/2}\tau)
=\frac{m^5_{\tilde{\tau}}}{48\pi m^2_{3/2}M^2_{\rm P}}
\left(1-\frac{m^2_{3/2}}{m^2_{\tilde{\tau}}}\right)^4\,,
\label{width}
\end{equation}
where the tau mass is omitted.

The lifetime of the NLSP $( \tau_{\rm NLSP} = 1/\Gamma_{\tilde{\tau}})$
could be long because the decay width is suppressed by $M^{-2}_{\rm
P}$. The lifetime is sensitive to the yet unknown gravitino mass. For
$\tilde{\tau}$ mass of 100~GeV, the lifetime of NLSP stau would be 60
nsec for a gravitino of $m_{3/2}=1$ keV ($ F= (2\times 10^6~{\rm
GeV})^2$), and $6\times 10^6$ sec for $m_{3/2}=10$~GeV($F=(6\times
10^9~{\rm GeV})^2$ ).    
If we can
determine the gravitino mass by measuring the energy of the emitted
tau lepton, the measurement of the stau lifetime
$(\Gamma_{\stau})^{-1}$ leads to a determination of a ``supergravity
Planck scale'', $M_{\rm P}[\mathrm{supergravity}] = M_{\rm
P}(\mstau,m_{3/2},\Gamma_{\stau})$ \cite{BHRY}. It would be a crucial
test of supergravity by comparing it with the macroscopically
determined Planck scale of Einstein gravity, $M_{\rm
P}[\mathrm{gravity}] = (8\pi G)^{-1/2}=2.43534(18)\times
10^{18}~\mathrm{GeV}$ \cite{PDG}. On the other hand, in the case where
the gravitino mass can not be directly determined due to limited
experimental resolutions, its mass, or SUSY breaking scale, can be
estimated from the lifetime in the assumption that Eq.~(\ref{width})
is correct.  Furthermore, a study of a rare 3-body decay, $\stau\to
\tau\,\gamma\,\psi_{3/2}$, can reveal the peculiar form of the
gravitino couplings, and may even determine the gravitino spin of
$3/2$ \cite{BHRY}.

Previously, the determination of the charged NLSP (CNLSP) lifetime
has been studied by
many authors and it was found that the CNLSP decay is detectable 
for $0.5{\rm m}<c\tau_{\rm CNLSP}<1{\rm km}$ \cite{short}.
In this paper, we study the possibility to measure the CNLSP
lifetime and  decay patterns  for the case when it is
long-lived so that the prompt decay inside the detector can be
neglected. 
For
the CNLSP with sufficiently slow velocity, it can be stopped in the
stopper material placed outside the main detector after losing its
energy from ionization in the main detector materials.\footnote{
For an earlier proposal to collect long-lived charged heavy fermions,
see Ref.~\cite{Goity:1993ih}.  Although details was not given, an idea
of placing a stopping material outside the main detector to collect
the CNLSPs has been pointed out in Ref.~\cite{FengCNLSPcollection}.}
Then the
lifetime of CNLSP thus stopped would be measured by detecting
subsequent decay of the CNLSP in the stopper. The stopper needs to be
a tracking detector to measure the stopping position and timing of the
parent CNLSP and also the momenta, positions and timings of the
charged particles and photons arising from the CNLSP decay. The
requirements of the stopper detector are also discussed.
Such study, if possible at future colliders such as the LHC, 
will also reveal the important interactions that may affect the big 
bang nucleosynthesis in the early universe.  

This paper is organized as follows. 
In section 2,  we discuss the requirements to study the CNLSP decay. 
In section 3, we discuss a possible study of the CNLSP at the 
LHC (ATLAS/CMS) \cite{unknown:1999fq, unknown:1999fr, cms}.  
The velocity distribution of CNLSP ($\beta_{\tilde{\tau}}$) 
at the LHC is shown at a sample 
point of the gauge mediation model \cite{gm}, which is studied in the 
ATLAS technical design report.  The number 
of CNLSPs stopped in a O(1) kton stopper is estimated, 
and it is found to be as large as 
O(5000) events for an integrated luminosity of
300~fb$^{-1}$ and squark masses of $m_{\tilde{q}}\sim 700$~GeV. 
So that, one can 
determine a lifetime of the  CNLSP 
of $\tau_{\rm CNLSP}<1000$ years or so, which corresponds to the 
SUSY breaking scale as large as O($10^{10}$)~GeV, 
or gravitino mass of O($100$)~GeV.
Here, we also discuss a possibility to make use of existing detector, 
taking Soudan 2 as an example.
In section 4, we will study the case at proposed future linear colliders \cite{LCs}. The
direct production of the CNLSP $\tilde{\tau}$ at a $e^+ e^-$ collider may
not be efficient because the cross section is suppressed by $(\beta_
{\tilde{\tau}})^3$, while one wants to keep $\beta_{\tilde{\tau}}$ 
relatively small so that CNLSP can be easily stopped.
When the mass difference
between $\tilde{e}$ and $\tilde{\tau}$ is small, an alternative option of $e^-e^-$
collision at the $\tilde{e}^-\tilde{e}^-$ threshold energy is more attractive, 
since the production cross section is only suppressed by
$\beta_{\tilde{e}}$. It is found that the number of  the CNLSPs stopped
in the stopper could be as large as $10^5$.  If this is the case,
one can study not only the lifetime of the CNLSP, but also 
the branching ratios and decay 
distributions of various CNLSP decay modes, including the 
rare decay modes. 

\section{Principle of Lifetime Measurements of Stopped CNLSPs}

We first discuss 
 how we could measure the lifetime of
the CNLSP, which can be produced at collider experiments.  The CNLSP would
be metastable and long-lived. Its lifetime $\tau_{\rm CNLSP}$ 
might range from O(100)
nsec to O(1000) years, depending on the CNLSP and gravitino masses.
When $c\tau_{\rm CNLSP}$ is too long to decay frequently in the detector ($\gsim$ 1000ns),
its lifetime cannot be measured directly at the main
detector of collider experiments. However the  CNLSPs with low
 velocity lose their kinetic energy by ionization loss, since they are
charged, and could be stopped at a massive stopper material, which can
possibly be placed outside the main detector.  A large number of the
CNLSPs thus stopped could be stored and accumulated in the stopper for a
significantly long time period. The CNLSPs then decay after $t\sim
\tau_{\rm CNLSP}$. 

The stopper should be an active detector with good tracking
capability to measure the lifetime of the CNLSP.
First of all, it should measure the ionization per 
unit length $-dE/dx$ to reject a $\mu$  track from that of  signal
CNLSP
\cite{Drees:1990yw}. 
The stopping timings $t_{\rm stop}$ and positions of all the CNLSP
candidates are measured and recorded one by one, as well as the 
CNLSP decay time $t_{\rm decay}$ and its decay products. The
lifetime of CNLSP is obtained from the distribution of a timing
difference $t_{\rm decay}-t_{\rm stop}$.  In particular, 
to correspond the parent CNLSP and
daughter decay products, the requirement that the decay products
originate from the stopping position of the parent CNLSP is of crucial
importance. Therefore, the stopper has to have 
a sufficiently high position resolution,
as well as a fine detector segmentation.

The stopper should be massive so that as large number of CNLSPs as
possible can be stopped, of the order of O(1) $\sim$ O(10)
kton. Typically, we need to have, at least, O(10) CNLSPs stop
and decay/year inside the stopper detector to estimate the lifetime in
accuracy of a few 10 \%. 
When a huge number of CNLSPs can be
accumulated, the distributions of the decay products can be studied.
The CNLSP is likely to be a $\tilde{\tau}$ and  it  
decays into a $\tau$ lepton. The $\tau$ further decays into 
$e^-$, $\mu^-$,  $\pi^-$ , $\rho^-$, $a_1$ and so on. $\rho^-$ and 
$a_1^-$ further decay into $\pi^-$ and   $\pi^0$('s), and momenta of the 
decay products  are parallel.  The decay distributions contain
physics information. 
The stopper detector might have good calorimetric performance for
them, then 
the gravitino mass 
can be calculated from the CNLSP mass and the tau energy, which is 
estimated from the end point of the tau jet energy distribution.
The composition of the $\tau$ jet 
is also important. The tau lepton is generally polarized depending 
on  the  left-right mixing of $\tilde{\tau}$, and 
the polarization of $\tau$ 
can be studied by measuring the $E_{\gamma}/E_{\rm jet}$ ratio \cite{nft}. 
In general, the $\pi^{\pm}$ is detected as 
an  ionizing track starting from the decay 
point and  it ends by interacting hadronically 
after penetrating   $\sim 80$ g /cm$^{2}$.  
On the other hand,  $\gamma$'s are converted into 
EM showers after one radiation length ( O(10)g/cm$^{2}$) .
Ideally, the detector needs to be segmented into 
cells with a thickness less than  O(10)g/cm$^{2}$ 
to any direction, 
so that $E_{\gamma}/E_{\rm jet}$ can be measured.

To estimate a suitable velocity ($\beta$) of the CNLSP to stop, 
a range $R$, which is a distance for a charged 
particle to penetrate before stopping,
is evaluated based on the Bethe-Bloch equation. 
For the value of the $\beta$ of interest,  the range $R$ 
is described by a function of  $\beta\times\gamma$ ($\gamma=1/\sqrt{1-\beta^2}$) 
with a mild dependence 
on types of the stopper material. For iron, 
$R/M=10$ g/(cm$^{2}\cdot$GeV) is necessary for a particle 
with  $\beta\gamma=0.43$, 
$R/M=30$ g/(cm$^{2}\cdot$GeV) for $\beta\gamma=0.62$, and
$R/M=50$ g/(cm$^{2}\cdot$GeV) for $\beta\gamma=0.75$ \cite{PDG}, where $M$ is the mass 
of a charged particle.
The CNLSP  with $\beta\gamma>1$  needs to go 
through  $>10000$ g/cm$^{2}$ when $m_{\rm CNLSP}=100$~GeV.  
It is difficult to stop such relativistic particles, 
therefore we will concentrate on the CNLSPs with slow velocity.

It is noted that most of the main detectors for collider experiments 
have significant thickness. Typically thickness 
of the detector is around 10 absorption length $\sim 1000{\rm g/cm^2}$
for pseudo-rapidity\footnote{
$\cos\theta = \tanh\eta$, where $\theta$ is the angle from the beam direction.} 
$|\eta|<1$ (see \cite{unknown:1999fq}).
They serve as a degrader to reduce kinetic 
energy of the CNLSPs before reaching the stopper. 
Some of the CNLSPs also stopped inside the detector, however, those 
stopped inside the detectors cannot be used to study their decay
unless the detectors are significantly modified.  While most of
those CNLSPs will stop in the calorimeter, the granularity of the
hadronic calorimeters might not be fine
and positions where they stopped
are not precisely determined.\footnote{For example, ATLAS hadronic calorimeter,
whose internal radius is 2.28 m, has position resolution $\Delta \eta
\times \Delta \phi=0.1\times 0.1$. In the radial direction, it is
segmented into
(only) three layers. }  It is then 
not obvious to reject  an activity in the calorimeter 
 from the cosmic ray events.
Furthermore, those detectors take data only at the time of
collisions. The decay of the particles with a lifetime longer than the
bunch spacing will not be tagged.

In the next subsequent sections, the velocity distribution of the CNLSPs
and the number of the CNLSPs stopped in the stopper are estimated, together
with accuracy of lifetime measurement, for the LHC and future linear
colliders separately.

\section{The CNLSP studies at high luminosity option of the LHC.}

At the LHC, the CNLSP can be produced from cascade decays of
squarks $\tilde{q}$  and gluinos $\tilde{g}$.  
The production cross section is sensitive to 
the squark and gluino masses. 
If $m_{\tilde{q}}\sim m_{\tilde{g}}\sim 700$~GeV (1TeV), $\sigma_{\rm SUSY}=20(3)$~pb
and $\sim 10^7$ ($10^6$) CNLSP will be produced for 
$\int dt {\cal L}=300$~fb$^{-1}$. For the case where $\tilde{q}$ and $\tilde{g}$ are 
as heavy as $\sim$ 2$\sim$3TeV, one may still search for the CNLSPs 
model independently from their direct productions which 
is discussed in \cite{Feng:2004mt}, however, the study of the 
CNLSP decays will be limited statistically. Therefore 
we concentrate on the case where CNLSPs are produced dominantly 
from $\tilde{q}$ and $\tilde{g}$  decays.

A case where the NLSP is a charged 
slepton is studied in paper \cite{Hinchliffe:1998ys}
at  a  gauge mediation model point  G2b\footnote{
The model point is outdated because the higgs mass 
is lighter than LEP II constraint. Taking $\tan\beta=20$ 
keeping other gauge mediation parameter, the 
higgs mass can be raised above the limit.
We study the G2b point, because mass resolutions have been 
investigated in detail at this point.  
The mass of sleptons are very degenerated for $\tan\beta=5$ so that 
$\tilde{e}\to\tilde{\tau} \tau e$ is closed and $\tilde{e}_R$ 
and $\tilde{\mu}_R$ are long-lived.
Therefore we show distribution for all sleptons in Fig.~1. } 
by using ISAJET \cite{Paige:2003mg}.  At this 
point, gluino decays dominantly into a squark and jet.  A squark 
decays dominantly  into gaugino like neutralinos $\tilde{\chi}^0_{1(2)}$ or  
 chargino $\tilde{\chi}^{\pm}_1$. 
The neutralinos and chargino further decay into lighter sleptons 
$\tilde{l}_R$;
\begin{equation}
(\tilde{g})\to \tilde{q} \to \tilde{\chi}^0_{1(2)}, \tilde{\chi}^+_1
\to \tilde{l}
\end{equation}
The relevant masses of the SUSY particles at G2b are as follows; 
$m_{\tilde{\chi}^0_1}= 112$~GeV, $m_{\tilde{\chi}^0_2}=203$~GeV,
$m_{\tilde{g}}=709$~GeV, $m_{\tilde{u}_L}=685$~GeV,
$m_{\tilde{u}_R} =664$~GeV, $m_{\tilde{e}_R}=102$~GeV, 
$m_{\tilde{\tau}}=101$~GeV.  More  details will be found in 
\cite{Hinchliffe:1998ys} and  \cite{unknown:1999fr}. 

The measurement of the slepton mass is performed
by using momentum measurement in inner tracker and 
time information at muon system. The mass error of the slepton 
would be around 0.1~GeV at this point \cite{Hinchliffe:1998ys}. 
This means that the gravitino mass $m_{3/2}$ is determined 
with the additional  $\tau_{\rm CNLSP}$ measurement
by using Eq.~(\ref{width}).

\begin{figure}[t!]
\includegraphics[width=8cm,angle=-90]{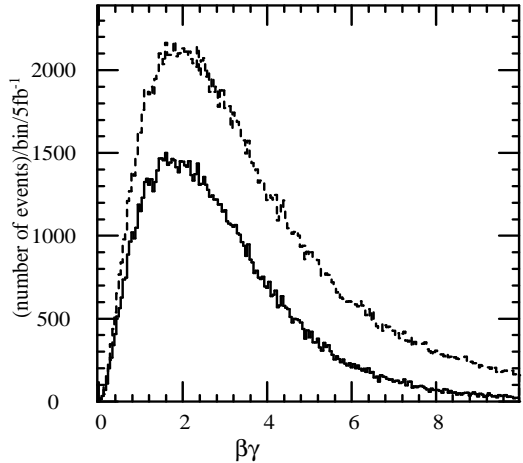}
\vskip -1cm
\caption{The  $\beta\gamma$ distribution of $\tilde{l}$ for 
$10^5$ SUSY events at the point G2b. The solid histogram  shows the 
distribution with $\vert \eta_{\tilde{l}} \vert <1$ and the 
dashed histogram  shows all $\tilde{l}$ distributions. 
 }
\label{betadis}  
\end{figure}

The typical transverse momentum 
of the CNLSP at the LHC  are determined 
by the mass difference between $\tilde{q}$ and $\tilde{\chi}^0_i$ 
or $\tilde{\chi}^{\pm}_i$. The  momentum of a $\tilde{\chi}$ 
from a stopped squark is expressed as  
\begin{equation}
\vert \vec{\rm p}^{\rm rest}_{\tilde{\chi}}\vert =\frac{m_{\tilde{q}}^2-m_{\tilde{\chi}}^2}{2m_{\tilde{q}}}.
\label{momentum}
\end{equation}
The $\tilde{q}$ production is dominated by the threshold 
production because the parton distribution is soft. Therefore 
the typical absolute momentum of $\tilde{\chi}$ peaks at 
$\vert{\rm p}^{\rm rest}\vert$, 
but it is boosted by parent $\tilde{q}$ velocity. The distribution 
therefore depends on the decay patterns and mass spectrum. 
The momentum of a $\tilde{l}$ from a  $\tilde{\chi}$ decay 
is similar to the $\tilde{\chi}$ momentum
when  the mass differences between 
$\tilde{\chi}$ and $\tilde{l}$ are small. 
To be quantitative, 
we generated $10^5$ SUSY events at the 
G2b point by using HERWIG \cite{Corcella:2000bw},
 which corresponds to $\int dt {\cal L}=
5{\rm fb}^{-1}$ at the LHC.   
In Fig.~\ref{betadis}, we show the $\beta\gamma$ distribution 
of $\tilde{l}$ at the G2b point. The dotted line shows 
the distribution of all events, while the 
solid line shows the event distribution with $\eta<1$.
The typical momentum is $\beta\gamma=\vert {\rm p}\vert/m\sim 2$, which 
is consistent with the estimate based on Eq.~(\ref{momentum}). From 
Fig.~\ref{betadis}, we can see  
the events are likely to be forward when $\beta\gamma$ is large,  
while the event distribution is more spherical when  $\beta\gamma\sim 0$.

\begin{table}[t!]
\begin{center}
\begin{tabular}{|c|c|c|}
\hline
CNLSP velocity& number of events& $\vert\eta\vert<1$\cr
\hline
$\beta\gamma<0.43$ & $1.68\times 10^3$ & $1.28\times 10^3$\cr
$0.43\leq\beta\gamma<0.62$& $2.85\times 10^3$& $2.14 \times 10^3$\cr
$0.62\leq\beta\gamma<0.75$&  $2.85\times 10^3$& $2.09 \times 10^3$\cr
\hline
\end{tabular}
\end{center}
\caption{Numbers of the CNLSP for different 
velocity regions at the 
point G2b for $ 10{^5}$ SUSY
events (5~fb$^{-1}$) at the LHC.}
\label{numbers}
\end{table}

In Table \ref{numbers}, the number of events for a few representative 
$\beta\gamma$ intervals for $10^5$ SUSY events  are shown. 
We  found that 
the 0.8\% of the total CNLSP (twice of the produced SUSY events) 
  has $\beta\gamma <0.43$. Because 
the mass of CNLSP is about 100~GeV for the  
G2b point, $R\sim 1000$g/cm$^2$ for
$\beta\gamma<0.43$ and they will be  stopped in the LHC detectors. 
We also found that 1.4\% of the total CNLSPs 
have $0.43\lsim\beta\gamma<0.62$
or  $0.62\lsim\beta\gamma<0.75$.

 The expected number of the stopped CNLSP
for a fixed stopper weight is inversely 
proportional to the square of distance of the stopper from the 
interaction point $R_{\rm IP}$. 
Note that  the ATLAS (CMS) detector has a cylindrical 
shape whose 
diameter is 20m (15m) respectively. 
To cover the ATLAS or CMS detector with 2000g/cm$^2$ stopper  
over the length of 
2 $R_{\rm IP}$ (corresponding to $\vert \eta \vert\lsim 1$) ,
we need 
more than $25\times(R_{\rm IP}/{\rm 10m})^2$  kton stoppers. 
Apparently $R_{\rm IP}$ should be larger than 
the half of the diameter at least. 

The expected  number of the trapped CNLSP  at the G2b point 
for integrated luminosity of  300~fb$^{-1}$
(3 year in high luminosity)
is given as a function of the stopper 
weight $M_T$ and  $R_{\rm IP}$
as 
\begin{equation}
N({\rm G2b})= 0.51\times 10^4 \left( \frac{M_T/1{\rm kton}}
{\left(R_{\rm IP}/{\rm 10m}\right)^2}\right)\,.
\label{NatG2b}
\end{equation}
Here we
have assumed that, if $M_T<25 (R_{\rm IP}/10{\rm m})^2$~kton, the
fraction of the solid angle of the stopper from the IP would be
reduced while keeping the thickness 2000g/cm$^2$.  

We now estimate the level of the background expected in O(1kton) stopper 
at the LHC. 
A possible source of the background is events from  
the charged current interaction  of  the atmospheric
neutrinos in the stopper. They are observed as charged particle
emissions contained in the stopper, just like the stopped 
CNLSP decays.

The background rate can be estimated by the event rate in 
Super Kamiokande. 
In \cite{Toshito:2001dk}, the number of multi GeV and 
partially contained events (with visible energy $E_{\rm vis}>1.33$~GeV) 
in Super Kamiokande  is reported as 
2256 events for 76~kton year.  On the other hand, 
the energy of leptons and jets are typically of 
the other of 10~GeV for the decaying CNLSP, 
and hence a significant  part of the atmospheric 
neutrino background will be removed by requiring  $E_{\rm vis}>10$~GeV
without losing too many signals.\footnote{High energy 
$\mu$s penetrate stoppers, and one needs magnet to measure the 
energies. 
One might have to veto $\mu$ like events to reduce 
the atmospheric backgrounds.}

The number of background events  may be calculated from the 
 neutrino energy distribution calculated
in \cite{Honda:2004yz}. The flux is roughly proportional to $1/E_\nu^3$ 
or less for $1$~GeV$<E_{\nu}<1000$~GeV, while the neutrino-interaction cross
section increases proportional to $E_{\nu}$.  The rate of
the charged current atmospheric neutrino events therefore 
scales as $1/E_{\nu}^2$. When $E_{\rm vis}>10$~GeV 
is required,  the rate would be reduced by a factor of $\sim 1/100$
compared to that for $E_{\rm vis}>1.33$~GeV, 
so that the number of  atmospheric neutrino
events would be less than 1/kton/year.  Furthermore,
if we record the  positions where
CNLSPs stopped, the number of background would  be reduced down to 
a negligible level.

Another source of background is the high energy neutrinos from 
$pp$ collisions. The heavy quarks $b$ and  $c$ are produced with 
very high rates, and their semi-leptonic 
decays produce high energy neutrinos. There are productions
of $W$ bosons as well,  and their decays produce neutrinos 
with $p_T\sim 45$~GeV.  The $\nu_{\mu}$ flux must be comparable to the muon flux at the LHC,  
as a  high energy $\nu_{\mu}$ is  always produced 
with a $\mu$ production. 

The 
production cross section of $\mu$ from various sources 
are plotted as the function of $p_T$ in Fig 2.1 in \cite{atlasmuontdr}. 
The dominant source of $\mu$ production 
is $b\to\mu$ for $p_T<35$~GeV and $d\sigma/dp_T(b\to\mu)=10^{-1}$$\mu$barn/GeV for 
$p_T=10$~GeV, $10^{-2}$$\mu$barn/GeV for $p_T=20$~GeV, and 
$10^{-3}$$\mu$barn/GeV for  $p_T=30$~GeV. For $p_T\sim 40$~GeV 
the dominant $\mu$ production comes from  $W\to\mu$ and 
$d\sigma/dp_T(W\to \mu)\sim  10^{-3}$$\mu$barn/GeV at the peak.  We assume the total 
neutrino flux is $10^{-1}$$\mu$barn for $p_T>20$~GeV
because we need to take into account $\nu_e$ and $\nu_{\tau}$ as well. 
Roughly  $3\times 10^{10}$ hard neutrinos are  produced for 
300~fb$^{-1}$. 
The charged current cross section of nucleon and neutrino
 is given in \cite{PDG} 
 as $\sigma/E_{\nu}\sim 0.677\times10^{-38}$~cm$^{-2}$.
The number of neutrinos which interact 
with 1 kton stopper at 10m from the IP is less than 0.1 
and can be neglected compared with atmospheric events. 

There might be other source of backgrounds such as 
neutrons from the surrounding environment and we need 
detailed studies to estimate the total background rate. The position 
measurements of the CNLSP will provide the 
good signal and background discriminations.

The number of CNLSPs decaying in the detector per year 
depends on the lifetime of the CNLSP. 
When $\tau_{\rm CNLSP}=100$ and $1000$ years, the number of the 
events decaying in the detector/year are 1\% and  0.1\% of 
the total number of events stopped, respectively. To have 10 CNLSP
decays in the 1 kton detector, the lifetime must be 
shorter than  1000 years for the G2b point when  1kton detector 
can be placed at $R_{\rm IP}=10$~m. 

When sparticle masses are higher than those of the G2b point, 
the number of stopped  CNLSPs  would be 
significantly smaller. First of all, the squark and 
gluino production cross sections decrease very quickly 
with increasing squark mass. 
At the G2b point where $m_{\tilde{g}}$ and $m_{\tilde{q}}\sim 700 $~GeV, the total
SUSY production cross section is calculated by HERWIG as  19.3~pb.
 On the other 
hand, at a model point SPS7 considered in Snowmass study 
\cite{Allanach:2002nj}, $m_{\tilde{g}}\sim 926$ GeV and the SUSY
production 
cross section is 5.2~pb. The CNLSP is $\tilde{\tau}$ with 
$m_{\tilde{\tau}}=120$ GeV for SPS7. The number of the CNLSPs with 
$0.43\leq\beta\gamma<0.62$ and $\vert \eta \vert <1$ 
is 2706  for $10^5$ produced SUSY 
particles. The stopper to stop all the  slepton with $\vert
\eta\vert<1$  and this  $\beta\gamma$ 
range must have a weight $\sim  30\times (R_{\rm IP}/10{\rm m})^2$ kton, 
because $m_{\tilde{\tau}}$  is larger. 
This means
\begin{equation}
N({\rm SPS7})= 0.14\times 10^4 \left( \frac{M_T/1 {\rm kton}}
{\left(R_{\rm IP}/{\rm 10m}\right)^2}\right). 
\end{equation} 
The number of stopped events is still large enough to measure
$\tilde{\tau}<100$ years, corresponding to $m_{3/2}\lsim 80$ GeV.

So far we have discussed the conditions to study CNLSP decay at 
the LHC. We now turn to the question if it is possible to 
construct a detector which satisfies these conditions. 
Here we take Soudan 2 detector as an example detector 
and consider if it can be operated close to the LHC detectors. 
The Soudan 2 is a fine-grained tracking calorimeter 
detector for proton decay searches \cite{soudan}.
The detector consists of  224 modules with each  
size 2.7m$\times$1m$\times$1m. 
A module then consists of drift tubes separated by 
14.7 mm between steel sheets. The steel sheets are 
source of protons for the experiment, but they could 
also work as the stoppers to stop the  CNLSPs. 
The total weight of the detector modules  is 0.96 kton.
The track resolution is 0.18 cm $\times$ 0.18cm $\times$ 1cm, 
which is small enough to isolate each stopped CNLSP.

There will be  a few Hz/cm$^2$ of 
charged particles at the muon system of 
the ATLAS detector at $\eta\sim 0$ \cite{atlasmuontdr}. 
The dead time of the drift tube of Sudan 2 is  100 $\mu$sec.
The number of muons which goes through a drift tube 
per the drift time (dead time)  
must be much less than one, so that the CNLSP tracks 
can be reconstructed.  
Each tube covers 147cm$^2$. The expected number of 
charged particles/sec/tube is therefore  
much less than 0.1/(100$\mu$sec), 
even at the surface of the LHC detectors.

The energy threshold of Soudan 2 detector is O(100) MeV for
both muons and electrons. The threshold  
is set for proton decay study. If new detectors can be 
built for the CNLSP study, 
the ratio (stopper mass)/(number of drift tubes)
may be increased 
while keeping most of the important aspects as the detector
for CNLSP decay products.

A more realistic 
study on the condition in the ATLAS/CMS cabin is needed 
to judge if it is a realistic option to study the 
CNLSP decays.  We stress that it is worth to investigate the
possibility to  start an experiment to 
stop and collect CNLSPs as early as possible, perhaps with the high luminosity operation of the 
LHC, if the CNLSPs  are accessible at the low luminosity run of 
the LHC and the CNLSP is metastable. 
 
\section{The CNLSP study at $e^-e^-$ linear collider}

As we have discussed in the previous sections, the
velocity of the stau, $\betastau$, should be small enough to be
stopped in the stopper. A crucial advantage for a linear collider (LC) is
that one can restrict the velocity of produced particles by adjusting
the beam energy.

At a $e^+e^-$ collider, the direct production of a $\stau^+\stau^-$
pair is suppressed near the threshold, since the cross section
decreases as $\sim \betastau^3$. On the other hand, at a $e^-e^-$
collider, a slepton pair of $\se^-\se^-$ is produced, with cross
section being proportional to a mere single power of the velocity,
$\betase$ \cite{Feng:1998ud}. 
Produced $\se^-$ decays into the NLSP $\stau$ with velocity
$\betastau$, which does not differ much from $\betase$ as long as the
mass difference is small. Therefore, in order to produce many low
velocity $\stau$s, the indirect production at a $e^-e^-$ collider may be
advantageous to the direct production at $e^+e^-$ collider. Hence, in
this section we investigate the possibility to study the CNLSP decay
at a $e^-e^-$ collider. A comparison between the $e^-e^-$ and $e^+e^-$
cases will be given at the end of this section.

The selectron $\se^-$ can decay into both of $\stau^-$ and $\stau^+$,
mainly via three body decays $\se^-\to \stau^- \tau^+ e^-$ and
$\se^-\to \stau^+ \tau^- e^-$, if these modes are kinematically
allowed.\footnote{Those three-body decays of selectrons were studied in
detail in Ref.~\cite{Ambrosanio:1997bq}.}  In the following, we
restrict ourselves to a pure ``right-handed'' NLSP stau
$\stau_R$ and right-handed selectron $\se_R$, 
for simplicity.  We
also assume that the modes $\se^-\to \stau^\pm \tau^\mp e^-$ are
kinematically allowed, i.e., $\mse-\mstau \gsim 1.78~\mathrm{GeV}$.
Then, for both the production and decay of the selectron, essentially
only Bino exchange contributes. For simplicity, we assume that the
lightest neutralino is a pure Bino, with mass $M_1 > \mse$.  Under
those assumptions, energy and angular distributions of indirectly
produced staus are determined only by 4 parameters, $\mstau$, $\mse$,
$M_1$, and $\betase$ (or equivalently the electron beam
energy). Generalizations to cases beyond those assumptions will be
straightforward.\footnote{If the modes $\se^-\to \stau^\pm \tau^\mp
e^-$ are kinematically forbidden, the main mode is either $\se^-\to
\stau^- \bar{\nu}_\tau \nu_e$ or $\se^-\to e^- \psi_{3/2}$ depending
on the parameters. If the former is the main mode, only $\stau^-$ will
be collected. If the latter is dominant, $\se^-$ plays the role of
NLSP and one can study the gravitino by its decay instead of $\stau$'s
decay.}

\begin{figure}[t!]
  \centerline{
    \scalebox{0.7}{\includegraphics{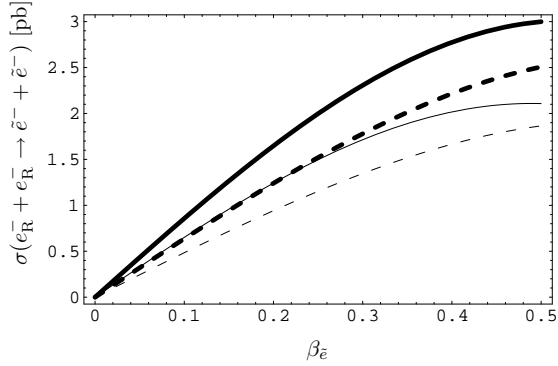}}
  } \caption{Cross section
    $\sigma(e^-_R + e^-_R\to
    \se^- + \se^-)$ for
    $\mse=170~\mathrm{GeV}$. We assume 100\%
    polarization for the $e^-$ beam.  $M_1 =
    180~\mathrm{GeV}$ (solid lines) and $M_1 =
    300~\mathrm{GeV}$ (dashed lines). The thick
    lines are the total cross section and the
    thin lines are for selectrons with
    $|\eta|<1$.}  \label{fig-sigma}
\end{figure}

\begin{figure}[t!]
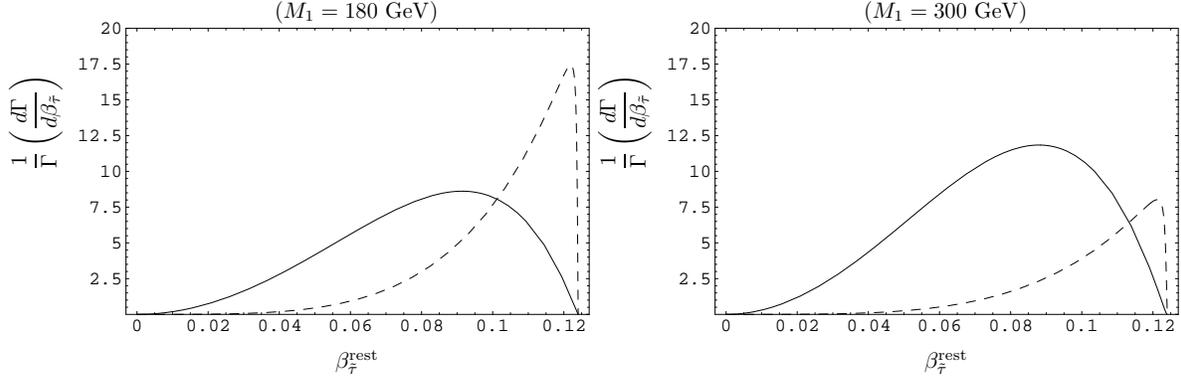

  \centerline{
    \scalebox{0.7}{\includegraphics{betastau-mB180.epsi}}
    \scalebox{0.7}{\includegraphics{betastau-mB300.epsi}}
  }
  \caption{Velocity distributions of $\stau^+$ (solid line) and
    $\stau^-$ (dashed line) at the rest frame of $\se^-$, for
    $\mse=170~\mathrm{GeV}$ and $\mstau=150~\mathrm{GeV}$. Bino mass
    is $M_1=180~\mathrm{GeV}$ (left) and
    $M_1=300~\mathrm{GeV}$ (right).}
  \label{fig-betastau}
\end{figure}

Production cross section of $\se^-\se^-$ pair for
$\mse=170~\mathrm{GeV}$ is shown in Fig.~\ref{fig-sigma}.\footnote{Here
we have neglected the beam effect and finite width
effects (cf. \cite{beamANDwidth}).}
The angular distribution of $\se^-$ is almost isotropic for small
$\betase$, and is mildly enhanced at beam direction for larger
$\betase$. Notice that all the produced selectrons have the same
velocity $\betase$ for a fixed beam energy. For instance, $\betase =
0.4$ corresponds to a center-of-mass energy
$E_\mathrm{cm}=371~\mathrm{GeV}$. Those selectrons decay into
$\stau^\pm$ via three body decays $\se^-\to\stau^\pm\tau^\mp
e^-$. Fig.~\ref{fig-betastau} shows the velocity distributions of the
produced $\stau^\pm$ at the rest frame of $\se^-$, for
$\mse=170~\mathrm{GeV}$ and $\mstau=150~\mathrm{GeV}$. In this frame,
$\stau^\pm$ are produced isotropically.

As can be seen in Fig.~\ref{fig-betastau}, there are interesting
differences between $\stau^+$ and $\stau^-$. First of all, the ratio
of $\stau^+/\stau^-$ events increases for larger
$M_1$ \cite{Ambrosanio:1997bq}. This is because the $\se^-\to \stau^+
\tau^- e^-$ mode picks up the Bino mass in the propagators (chirality
flipped), while the other mode does not.  Secondly, one can see that
$\stau^-$ distribution has a peak at larger velocity than $\stau^+$.

\begin{figure}[t!]
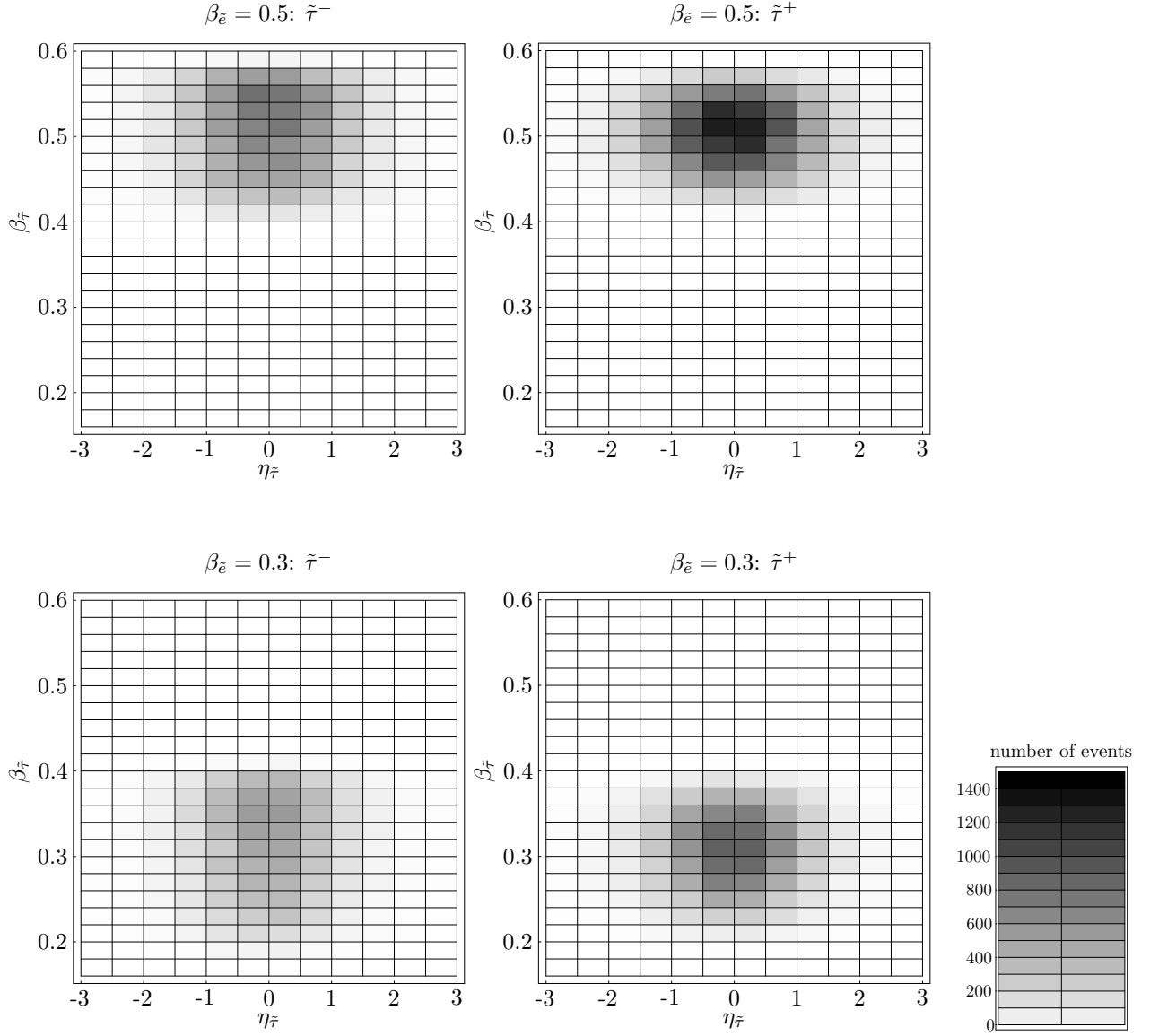

  \leftline{
    \scalebox{0.65}{\includegraphics{Stau05.epsi}}
    \scalebox{0.65}{\includegraphics{Astau05.epsi}}
  }
  \vspace{1cm}
  \leftline{
    \scalebox{0.65}{\includegraphics{Stau03.epsi}}
    \scalebox{0.65}{\includegraphics{Astau03.epsi}}
    ~
    \scalebox{0.4}{\includegraphics{Legend.epsi}}
  }
  \caption{($\eta$, $\beta$) distributions of $\stau^\pm$ at the
    laboratory frame, for $\mstau=150~\mathrm{GeV}$,
    $\mse=170~\mathrm{GeV}$, and $M_1=180~\mathrm{GeV}$. The
    velocity of the selectron is $\betase=0.5$ for upper figures,
    and $\betase=0.3$ for lower figures. Total number of events are
    normalized so that it corresponds to integrated luminosity
    $10~\mathrm{fb}^{-1}$.}
  \label{fig-etabeta}
\end{figure}

Now we can calculate the ($\eta$, $\beta$) distributions of
$\stau^\pm$ at the laboratory frame by combining the angular
distribution of $\se^-$ with the energy distributions of $\stau^\pm$
as shown in Fig.~\ref{fig-betastau}. The results are shown in
Fig.~\ref{fig-etabeta}, for $\mstau=150~\mathrm{GeV}$,
$\mse=170~\mathrm{GeV}$, and $M_1=180~\mathrm{GeV}$. Here, we have
normalized the total number of events so that it corresponds to
integrated luminosity $10~\mathrm{fb}^{-1}$.
One can see that the $\stau^+$ distribution has a sharper peak at
$\betastau\simeq \betase$ than $\stau^-$. This is because
$\stau^+$ has typically smaller velocity than $\stau^-$ at the rest
frame of $\se^-$, as can be seen in
Fig.~\ref{fig-betastau}.

Finally, we estimate the total number of $\stau^\pm$s which are stopped at
a material. Here, as examples, we assume $1000~\mathrm{g/cm^2}$,
$3000~\mathrm{g/cm^2}$, and $5000~\mathrm{g/cm^2}$ of iron placed at
the region $|\eta| < 1$. For
$\mstau=150~\mathrm{GeV}$, they can stop staus with velocities
$\betastau\lsim 0.37$, $\betastau\lsim 0.48$, and $\betastau\lsim
0.54$, respectively \cite{PDG}. We have counted the number of events in these
regions, for the parameter set used in Fig.~\ref{fig-etabeta}. The
results are shown in Table.\ref{table-final}.
\begin{table}[t!]
 \begin{center}
  \begin{tabular}{|c|rr|rr|rr|rr|}
    \hline
    & \multicolumn{2}{|c|}{total}
    & \multicolumn{2}{|c|}{$1000~\mathrm{g/cm^2}$ }
    & \multicolumn{2}{|c|}{$3000~\mathrm{g/cm^2}$ }
    & \multicolumn{2}{|c|}{$5000~\mathrm{g/cm^2}$ }
    \\ 
    & $\times 10^4$ 
    & $\times 10^4$ 
    & $\times 10^4$ 
    & $\times 10^4$ 
    & $\times 10^4$ 
    & $\times 10^4$ 
    & $\times 10^4$ 
    & $\times 10^4$ 
    \\ \hline
    $\betase=0.2$
    & 1.47 $\stau^-$
    & 1.83 $\stau^+$
    & 1.11 $\stau^-$
    & 1.38 $\stau^+$
    & 1.11 $\stau^-$
    & 1.38 $\stau^+$
    & 1.11 $\stau^-$
    & 1.38 $\stau^+$
    \\
    $\betase=0.3$
    & 2.06 $\stau^-$
    & 2.56 $\stau^+$
    & 1.30 $\stau^-$
    & 1.78 $\stau^+$
    & 1.54 $\stau^-$
    & 1.90 $\stau^+$
    & 1.54 $\stau^-$
    & 1.90 $\stau^+$
    \\
    $\betase=0.4$
    & 2.47 $\stau^-$
    & 3.08 $\stau^+$
    & 0.49 $\stau^-$
    & 0.48 $\stau^+$
    & 1.71 $\stau^-$
    & 2.22 $\stau^+$
    & 1.81 $\stau^-$
    & 2.25 $\stau^+$
    \\
    $\betase=0.5$
    & 2.67 $\stau^-$
    & 3.33 $\stau^+$
    & 0 $\stau^-$
    & 0 $\stau^+$
    & 0.60 $\stau^-$
    & 0.65 $\stau^+$
    & 1.38 $\stau^-$
    & 1.95 $\stau^+$
    \\    \hline
  \end{tabular}
 \end{center}
 \caption{Estimated numbers of $\stau^\pm$ stopped in the material at
 $e^-e^-$ collider with integrated luminosity $10~\mathrm{fb}^{-1}$.
 As a stopping material, we have assumed iron placed at $|\eta|<1$
 region. Parameters are $\mstau=150~\mathrm{GeV}$,
 $\mse=170~\mathrm{GeV}$, and $M_1=180~\mathrm{GeV}$.}
 \label{table-final}
\end{table}

Given the weight of stopping material, 
one can tune the beam energy,
thereby the selectron velocity $\betase$, 
to maximize the obtained
number of staus.
We again assume that the total thickness which particle has to go through
within the LC detector is about 1000g/cm$^2$.  The optimal velocity is
$\beta_{\tilde{e}}\simeq 0.4$ for the stopper of the thickness 2000
g/cm$^2$, when $3.0\times 10^4$ CNLSPs can be stopped in the stopper
of $25 {\rm kton}\times(R_{\rm IP}/{\rm 10m})^2$ for 10~fb$^{-1}$.
For integrated luminosity $100~\mathrm{fb}^{-1}$ (corresponding to 250
days for $e^-e^-$ luminosity in \cite{Brinkmann:2001qn})
 and a 10kton
stopper, the total number of the CNLSP is therefore
\begin{eqnarray}
  N = 1.2\times 10^5\left( \frac{M_T/10 {\rm kton}}
{\left(R_{\rm IP}/{\rm 10m}\right)^2}\right), 
\end{eqnarray}
for the present parameter set. 

To illustrate the model dependence, let us consider another example,
$(m_{\tilde{\tau}},m_{\tilde{e}},M_1)=(100,103,110)$ GeV,
where the masses are close to these for G2b point. 
This case is
better than the previous one in several aspects. First of all, smaller
$m_{\tilde{e}}$ and $M_1$ lead to a larger production cross section of
$\tilde{e}$. Secondly, smaller $m_{\tilde{\tau}}$ means shorter range
$R$ for a fixed velocity $\beta$. Finally, because of the small mass
difference between $m_{\tilde{e}}$ and $m_{\tilde{\tau}}$,
$\tilde{\tau}$'s velocity has a very narrow distribution,
$|\beta_{\tilde{\tau}}-\beta_{\tilde{e}}|\lsim
\beta_{\tilde{\tau}}^{\rm rest}\simeq 0.024$. Assuming again
$1000~\mathrm{g/cm^2}$ thickness of the LC detector at
$|\eta|<1$, a $1000~\mathrm{g/cm^2}$ stopper will be sufficient to stop
CNLSPs with velocity $0.4\lsim \beta_{\tilde{\tau}} \lsim
0.48$. Therefore, for $\beta_{\tilde{e}}=0.44$, all the generated
$\tilde{\tau}$s ($0.42<\beta_{\tilde{\tau}}<0.46$) at $|\eta|<1$ will
be stopped with a 12.6$\times (R_{\rm IP}/10{\rm m})^2$kton
stopper. We found that the total number of collected $\tilde{\tau}$ is
\begin{eqnarray}
  N = 0.9\times 10^6\left( \frac{M_T/10 {\rm kton}}
{\left(R_{\rm IP}/{\rm 10m}\right)^2}\right), 
\end{eqnarray}
for integrated luminosity $100~\mathrm{fb}^{-1}$. Note that we have
assumed a $1000~\mathrm{g/cm^2}$ stopper, and hence a wider area is
covered for a fixed weight $M_T$ than the previous cases. For a 1kton
stopper, roughly ten times more CNLSPs can be collected than
Eq.(\ref{NatG2b}).

So far we have assumed the thickness of the LC detector is
uniform for $|\eta|\lsim 1$. Ideally, given the masses
$m_{\tilde{e}}$ and $m_{\tilde{\tau}}$, the total weight of the
stopper $M_T$ and the profile of the collider detector, one can
optimize the shape (thickness) of the stopper and the electron beam
energy in order to maximize the collected number of CNLSPs.

With the above number of stopped CNLSP, one can measure the SUSY breaking scale
from the lifetime very precisely. 
In addition, the mass of gravitino
would be measured directly  from the measurement of
the upper end  of the $\tau$ jet energy from the CNLSP decays. 
The end point is 
expressed as $E_{\tau}\sim
(m^2_{\tilde{\tau}}-m^2_{3/2})/(2m_{\tilde{\tau}})$.  
If the error of energy
scale of the $\tau$ jet $\Delta E/E$ is 
$\epsilon$,  a gravitino mass of $m_{3/2}\gsim \sqrt{\epsilon}$
$\times m_{\rm NLSP}$  can be measured.    For example, 
for $m_{\tilde{\tau}}=150$~GeV and
$m_{3/2}=30$~GeV~\footnote{These masses are (marginally) consistent 
with the BBN constraint on the NLSP decay \cite{Feng:2004mt}.}, 
the tau energy is 72~GeV. If it is measured within
$\epsilon=3\%$, gravitino mass will be determined as $m_{3/2} =$
(16--39)~GeV. The lifetime $\tau_{\rm CNLSP}=95$ days 
and the mass $m_{\tilde{\tau}}$ will be measured much more accurately.
Notably this  leads to a measurement 
of Planck scale  $M_{\rm P}=$ (1.9--4.6)$\times
10^{18}$~GeV by using Eq.~(\ref{width}).

We found that there might be O$(10^5)$ events available for the CNLSP
study if $e^-e^-$ linear collider can produce $\tilde{e}^-\tilde{e}^-$ near the
threshold and $\tilde{e}^-$ decays into a CNLSP which is close to
$\tilde{e}^-$ in mass.  When the lifetime of the gravitino is order of
a year or less, all of the stopped CNLSP decay will be observed, and
about 1\% of them decay through its rare decay mode such as
$\tilde{\tau}\rightarrow\gamma \tau \psi_{3/2}$.

In paper \cite{BHRY}, the importance to study such rare decay modes
is discussed. The decay into gravitino is induced by the interaction

\begin{equation} L_{3/2}=-\frac{1}{\sqrt{2}M_{\rm P}} \left[(D_{\nu}
\tilde{\tau}_R)^{*} \bar{\psi}^{\mu} \gamma^{\nu}\gamma_{\mu} P_R \tau +
(D_{\nu}\tilde{\tau}_R ) \bar{\tau}P_L
\gamma_{\mu}\gamma^{\nu}\psi^{\mu} \right] + (L \leftrightarrow R)\,.
\end{equation} 
In the limit of very large gaugino masses, 
the 3 body
decay proceeds through the photon emission from the any of the charged
external lines of the decay $\tilde{\tau}\to \psi_{3/2}\tau$,
or the 4 point photon-gravitino-tau-stau interactions.

The peculiarity of the stau decay into the gravitino compared to 
those by ordinary
Yukawa type interaction may be seen more clearly in the limit
$m_{\tilde{\tau}}\gg m_{3/2}$ when the helicity 1/2 component
dominates the gravitino interaction, because the wave function of
$h=1/2$ components will be enhanced proportional to its energy. Indeed
it is reasonable to assume this limit for the rare mode study, because
the CNLSP lifetime should not be too much larger than O(year) to make
a full use of the all stopped CNLSPs.  This means the gravitino mass
cannot be comparable to the stau mass, in that case lifetime easily
exceeds O(year).

In the limit $m_{\tilde{\tau}}\gg m_{3/2}$, the $\tilde{\tau}$
decay then may be governed by the effective interaction of goldstino
$\chi$ where $\psi_{\mu}\sim \sqrt{2/3} (\partial_{\mu}\chi)/ m_{3/2}$,
\begin{equation} L_{\rm eff}= \frac{m_{\tilde{\tau}}^2}{\sqrt{3} M_{\rm P}
m_{3/2}} \left( \tilde{\tau}^*_R \bar{\chi} P_R \tau + \tilde{\tau}_R
\bar{\tau} P_L \chi \right)+(L\leftrightarrow R) -
\frac{m_{\tilde{\gamma}}}{4\sqrt{6} M_{\rm P}m_{3/2}}
\bar{\chi}\left[\gamma^{\mu},\gamma^{\nu}\right]
\tilde{\gamma}F_{\mu\nu}.  \label{eff}
\end{equation} 
The first term is Yukawa type interaction and
renormalizable, while the last term (photon-photino-gravitino
interaction) is a non-renormalizable one.  The photon-photino-gravitino
interaction contributes to the CNLSP decay through the photino
exchange.  The last term manifests itself in the three body decay
distribution of gravitinos, and may be extracted from the 
study of the decay distribution. 

The difference between the decay pattern of the gravitino to some
fermion, which comes from a non-renormalizable terms, may be studied
by measuring the photon energy and angle between photon and
$\tau$ \cite{BHRY}.  For the type of the drift tube detector
discussed in the previous section, the angle between the tau jets and
photon would be measured precisely.  The energy resolution, on the
other hand, would be around 10\% event by event. The statistical
significance of the measurement of the effective couplings in
Eq.~(\ref{eff}) at a future linear collider will be studied elsewhere. 

Before closing this section, let us compare the cases of $e^+e^-$ and
$e^-e^-$ colliders. As noted before, the production cross sections
$\sigma(e^+e^-\to \tilde{l}_R^+\tilde{l}_R^-)$ are suppressed by
$\beta^2$ compared to $\sigma(e^-e^-\to \tilde{e}^-_R\tilde{e}^-_R)$,
where $\tilde{l}_R = \tilde{\tau}_R$, $\tilde{\mu}_R$,
$\tilde{e}_R$.  Quantitatively, they are even more suppressed. We have
numerically checked that 
$\sigma(e^+e_R^-\to \tilde{\tau}_R^+\tilde{\tau}_R^-)
/\sigma(e_R^-e_R^-\to \tilde{e}^-_R\tilde{e}^-_R) < 0.015$ 
and
$\sigma(e^+e_R^-\to \tilde{e}_R^+\tilde{e}_R^-)
/\sigma(e_R^-e_R^-\to \tilde{e}^-_R\tilde{e}^-_R) < 0.014$ 
for $\beta<0.5$, 
for both of the above examples 
$(m_{\tilde{\tau}},m_{\tilde{e}},M_1)=(150,170,180)~\mathrm{GeV}$ and 
$(100,103,110)~\mathrm{GeV}$.
Even for unpolarized electron beam, the ratios are less than 0.037.
On the other hand, the
luminosity of a $e^-e^-$ collider becomes smaller than $e^+e^-$ due to
the de-focusing effect of the same charge beams. It is found in
Ref.~\cite{Brinkmann:2001qn} that, assuming identical beam parameters,
the luminosity of $e^-e^-$ mode is seven times smaller than the
$e^+e^-$ mode. Therefore, $e^-e^-$ mode can produce at least four
times more  low velocity CNLSP staus than $e^+e^-$, per
time.

Another possibility is a $\tilde{e}^\pm_R\tilde{e}_L^\mp$ pair
production at $e^+e^-$ collider, at the threshold $E_{\rm cm}\simeq
m_{\tilde{e}_R}+m_{\tilde{e}_L}$.\footnote{We thank
P.~M.~Zerwas for pointing out this option.} Although
$\sigma(e^+e^-\to \tilde{e}^+_R\tilde{e}^-_R)$ is suppressed by
$\beta^3$, $\sigma(e^+e^-\to \tilde{e}^\pm_R\tilde{e}^\mp_L)$ is
suppressed only by $\beta$.
In the limit of 
$m_{\tilde{e}_L}\simeq m_{\tilde{e}_R}$ 
and 
$\beta\ll 1$, 
$\sigma(e_R^+e_R^-\to \tilde{e}_L^+\tilde{e}_R^-)
=
\sigma(e_L^+e_L^-\to \tilde{e}_R^+\tilde{e}_L^-)$
is eight times smaller than 
$\sigma(e_R^-e_R^-\to \tilde{e}_R^-\tilde{e}_R^-)$,
because of the bino coupling ($\times 1/4$),
absence of $u$-channel ($\times 1/4$) and
non-identical final states ($\times 2$).
This factor offsets the advantage of the luminosity.
Besides, $\tilde{e}_L$ is typically heavier than $\tilde{e}_R$
by O(1) factor. This leads to additional suppression of the
cross section $\sigma\propto M_1^2/E_{\rm cm}^4$ by a factor of
$(2m_{\tilde{e}_R} / (m_{\tilde{e}_L} + m_{\tilde{e}_R}) )^4$.
A CNLSP generated by $\tilde{e}_L$ is too rapid to stop in the stopper,
resulting in another suppression of factor $1/2$.
Therefore, the $e^-e^-$ option is still likely 
advantageous to the $e^+e^-$ option.

\section{Discussion}
In this paper we have discussed the possibility to study the 
charged NLSP (CNLSP) $\tilde{l}$ decaying into the LSP gravitino 
at future collider experiments such 
as the LHC or $e^+e^-$/$e^-e^-$  linear colliders. 
O$(10^4)$ (LHC)  to O$(10^5)$ (LC) CNLSP  
may be stopped if  a 1 kton stopper  
is placed close to the detectors at the interaction point  
for the case where sparticle masses are 
close to the current experimental 
limits. 

A CNLSP lifetime up to $10^3$ years will be accessible
when $10^4$ CNLSPs are  stopped in the stopper, because more than 10 CNLSPs 
decay inside the detector/year in that case. 
If the CNLSP production rate is low and their lifetime is too long, the signal
to background ratio could be problematical. In such a serious case, we
might be forced to do detection only during beam-off period, with cost
of lower detection efficiency. Detail estimation of the signal to
background ratio needs further simulations with  realistic 
detector design and experimental situations.

For  O$(10^5)$ CNLSPs, one 
can study the rare decays of the CNLSP. It should be noted that 
not only the gravitino-CNLSP interaction, but also all the other 
exotic interactions involving CNLSP can be investigated by collecting 
metastable CNLSP.

In this paper we restrict ourselves to the case where the long-lived
charged particle is the CNLSP $\tilde{\tau}$. Stable gluino scenario
\cite{Chen:1996ap,Baer:1998pg} or split SUSY scenario where the gluino
lifetime is long because of superheavy sfermions
\cite{Arkani-Hamed:2004fb} predict a hadronic charged stable or
meta-stable particle.
They may also be stopped
by a massive stopper detector. In the case of split SUSY scenario, the
gluino lifetime depends on the superheavy sfermion mass scale.  However, the
number of stopped gluinos would be significantly smaller than those
expected for the case where $m_{\rm CNLSP}\sim 100$~GeV studied in
this paper, because the stopping range increases proportional to the
masses.

The physics that can be extracted from the lifetime measurement is
rich. 
One can determine the scale of the hidden sector SUSY breaking by
measuring the NLSP lifetime $\tau_{\rm NLSP}$.  This restricts the
possible SUSY breaking scenario strongly.  The direct information on
the NLSP lifetime also constrains the possible cosmological
scenario.
If the NLSP lifetime is longer than O$(1)$ sec, the decay
of NLSP may change the abundances of the light elements of nuclei in
the early Universe, so that the upper limit on
the number density at the time of NLSP decay in the early Universe
would be obtained \cite{recentBBN}. 
This in turn severely restricts the parameters of supersymmetric models 
\cite{Feng:2004mt,gravLSPinCMSSM}.

If the CNLSP turns out
to be long-lived, high priority must be 
given to the study of the CNLSP
lifetime so that the measurement can be done during the high
luminosity run of the LHC.
Suitable 1kton detectors exist already for proton 
decay searches/neutrino detector now, and we discussed  
the Soudan 2 detector as an example.  It is  important  
to keep these detectors after their original physics 
target is achieved by the time when the physics results 
at the low luminosity run is obtained at the LHC. 

\subsection*{Note added}
After this paper appeared in e-print archive, Feng and Smith also
submitted a paper~\cite{Feng-Smith} on the number of slepton that
would be trapped at the stopper near the LHC and LC. They consider the
possibility to place a water tank outside the main detector, then drain
the water to a reservoir in low background environment.  This perhaps is
not useful for the NLSP with lifetime much shorter than draining
period.

\subsection*{acknowledgment}
We thank A.~Ibarra, M.~Ratz, F.~D.~Steffen, S.~Tanaka
and P.~M.~Zerwas for valuable discussions.
K.H. thanks J.~L.~Feng and M.~Sher for the exciting conversation 
at the SUSY'04 conference.
The
authors thank the Yukawa Institute for Theoretical Physics at Kyoto
University, where this work was initiated during the YITP-W-04-04
workshop on "Progress in Particle Physics".  This work is supported in
part by the Grant-in-Aid for Science Research, Ministry of Education,
Science and Culture, Japan (No.14540260 and 14046210 for
M.M.N.). M.M.N. is also supported in part by a Grant-in-Aid for the
21st Century COE ``Center for Diversity and Universality in Physics''.

\end{document}